\documentclass[aps,prl,reprint,superscriptaddress]{revtex4-1}
\usepackage{hyperref} 
\usepackage{graphicx}
\usepackage{amsmath}
\newcommand{\eq}[1]{\begin{equation}#1\end{equation}} 
\begin{document}


\title{Crossover between electron-phonon and boundary resistance limited thermal relaxation in copper films}


\author{L. B. Wang}
\email[]{libin.wang@aalto.fi}
\affiliation{QTF Centre of Excellence, Department of Applied Physics, Aalto University, FI-00076 Aalto, Finland}
\author{O.-P. Saira}
\affiliation{Department of Physics and Kavli Nanoscience Institute, California Institute of Technolog, Pasadena, CA, USA}
\affiliation{Computational Science Initiative, Brookhaven National Laboratory, Upton, NY 11973}
\author{D. S. Golubev}
\affiliation{QTF Centre of Excellence, Department of Applied Physics, Aalto University, FI-00076 Aalto, Finland}
\author{J. P. Pekola}
\affiliation{QTF Centre of Excellence, Department of Applied Physics, Aalto University, FI-00076 Aalto, Finland}

\date{\today}

\begin{abstract}
We observe a crossover from electron-phonon (ep) coupling limited energy relaxation to that governed by thermal boundary resistance (pp) in copper films at sub-kelvin temperatures. Our measurement yields a quantitative picture of heat currents, in terms of temperature dependences and magnitudes, in both ep and pp limited regimes, respectively. We show that by adding a third layer in between the copper film and the substrate, the thermal boundary resistance is increased fourfold, consistent with an assumed series connection of thermal resistances.
\end{abstract}


\maketitle
Investigation of energy relaxation of electrons in normal metal films is important for understanding the underlying physics as well as for applications \cite{Kanskar1993,Schmidt2004,Giazotto2006,Pinsolle2016}. Especially for mesoscopic devices at low temperature, where the dominant thermal wavelength ${\lambda}$ is comparable to the device dimension, phonons in the films could be two-dimensional (2D), and it has been shown both experimentally and theoretically that the reduced phonon dimension does affect the energy relaxation of electrons in thin films \cite{Nabity1991,Glavin2002,Qu2005,Karvonen2007}. Heat transport by phonons, electrons, and photons has been studied experimentally in mesoscopic devices \cite{Schwab2000,Chiatti2006,Meschke2006,Jezouin2013,Partanen2016,Tavakoli2018}. From the application point of view, a good understanding of energy relaxation in metal films is important, e.g., for calorimetry and bolometry \cite{Pekola2013}. Decreasing the heat conductance from the metal film absorber to the environment will enhance the energy resolution, but on the other hand, it makes the device slower. For a transition-edge sensor, unaccounted-for thermal boundaries can affect the noise and energy resolution \cite{Kinnunen2012}. Finally, for a normal-metal/insulator/superconductor (NIS) junction cooler, quick thermalization of the secondary electrode is favorable in order to increase the cooling efficiency \cite{Lowell2013}.

In a heated normal metal film on a dielectric substrate electrons within the film relax by electron-electron (ee) interactions, and the energy is dissipated to the environment mainly by electron-phonon (ep) coupling to the film phonons, which is characterized by ep thermal coupling resistance $R_{ep}$. Film phonons are coupled to the substrate phonons, which are usually considered to constitute the heat bath for the device, by phononic coupling. The corresponding thermal resistance between phonons in the film and the substrate is the thermal boundary resistance $R_{pp}$. If the ee interactions are assumed to be much faster than other processes, the energy relaxation of the electrons in the film is determined by $R_{ep}$ and $R_{pp}$, with the weaker of the two governing the energy relaxation process. For thin films at low temperatures, ep coupling strength is weak and it becomes the bottleneck of the energy relaxation. With increasing the temperature or film thickness the ep coupling gets realtively stronger and the heat transport across the boundary between the film and the substrate becomes the limitation for the energy relaxation.

Electron-phonon coupling in metal films at low temperatures has been actively studied during the last decades. In particular, the effect of disorder and phonon dimensionality on the ep coupling strength have been intensively discussed \cite{DiTusa1992,Qu2005,Sergeev2000,Underwood2011}. Thermal boundary resistance between metals and dielectric substrates  has also been well investigated. Experimental observations can be explained with either Acoustic Mismatch Model (AMM) or Diffuse Mismatch Model (DMM) \cite{Swartz1989}. AMM describes phonon heat transfer through a flat interface between perfect crystals. In analogy to the Snell's law for the electromagetic waves, only the phonons with the incident angles below the critical one are transmitted through the interface. The critical angle is determined by the acoustic properties of the materials on both sides of the boundary. DMM assumes diffusive phonon scattering at the interface, and hence the phonon transmission probability depends only on the phonon densities of the states and sound velocities on both sides. In the case of solid-solid boundaries the mismatch in sound velocities and phonon mode densities is usually small, and the two models give similar predictions.

Here, we will present the experimental results showing the crossover between ep and boundary resistance limited thermal relaxation in Cu films at sub-kelvin temperatures. For Cu film with 50~nm in thickness, we found the energy relaxation to be limited by ep coupling in the full temperature range explored. By increasing the film thickness to 300~nm,  the thermal boundary resistance limits the energy relaxation, and we are able to quantify the heat transport between the metal/substrate interface directly from the experiments. By adding a third thin layer of film between the Cu film and the substrate, the thermal boundary resistance is increased fourfold, consistent with the assumption of a series connection of the thermal boundary resistances.

\begin{figure}[t]
	\centering
	\includegraphics[width=0.5\textwidth]{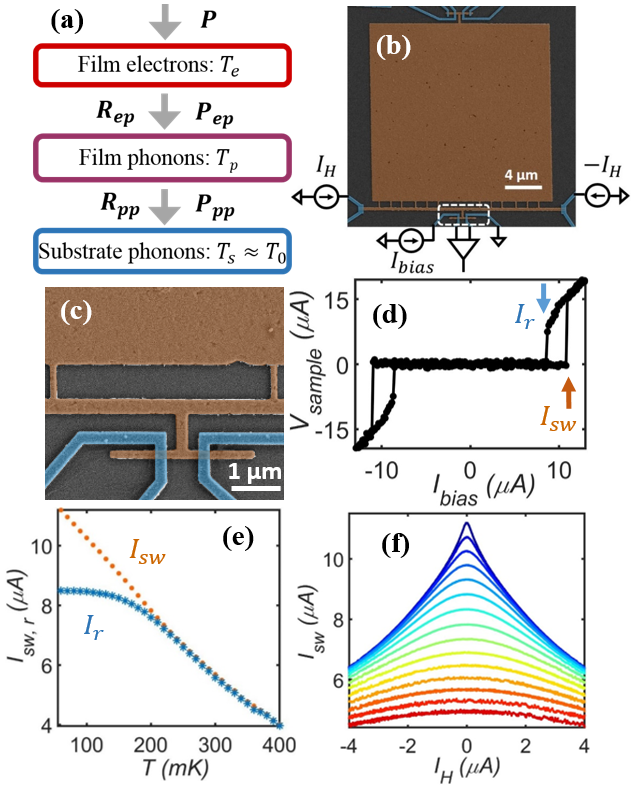}
	\caption{(a) The thermal model for the energy flow for a heated metal film on a substrate. (b) False-color SEM image of a sample together with the measurement setup. (c) Zoom-in of the rectangular area within the white dashed line in (b) showing the JJ thermometer connected to the heater and to the large Cu pad (Al: blue, Cu: brown). (d) IV curve of the JJ thermometer. (e) $I_{sw}$ and $I_{r}$ as a function of the bath temperature without heating applied to the Cu film. This equilibrium temperature dependence of $I_{sw}$ is used as the calibration for the JJ thermometer. (f) $I_{sw}$ as a function of $I_{H}$ for bath temperatures from 60~mK (blue) to 340 mK (red) with 20~mK intervals. The decrease of $I_{sw}$ when passing $I_{H}$ through the Cu film indicates heating of the electrons.}
	\label{fig1}
\end{figure}

For a heated metal film on a substrate, the energy flow is shown in the thermal model in Fig.~\ref{fig1}(a). Within the film, the energy flow rate from electrons to phonons is described by
\eq{
	P_{ep} = {\Sigma}V(T_{e}^{n}-T_{p}^{n}). \label{eq1}
}
Here, $T_e$ and $T_p$ are the electron and phonon temperatures in the film, $V$ is the metal volume, $n = 5$ for clean normal metals and $\Sigma$ is the material-specific ep coupling constant \cite{Wellstood1994}. The coupling between film phonons and the substrate phonons is characterized by
\eq{
	P_{pp} = kA(T_{p}^{4}-T_{s}^{4}), \label{eq2}
}
where $T_s$ is the substrate phonon temperature, $A$ is the contact area, and $k$ is the interface-material-dependent constant which can be calculated with DMM as
 \eq{
	k = \frac{\pi ^2}{120}\frac{k_B^4}{\hbar ^3}\frac{(\frac{1}{c_{1L}^2}+\frac{2}{c_{1T}^2})(\frac{1}{c_{2L}^2}+\frac{2}{c_{2T}^2})}{\frac{1}{c_{1L}^2} + \frac{2}{c_{1T}^2} + \frac{1}{c_{2L}^2} + \frac{2}{c_{2T}^2}}. \label{eq4}
}
Here, $c_{xL}$ and $c_{xT}$ are the speed of longitudinal and transverse sound on the side $x$ of the interface. For small temperature differences, the ep thermal coupling resistance is expressed as $R_{ep} = 1/5\Sigma VT^4$, and the thermal boundary resistance as $R_{pp} = 1/4kAT^3$. $T_s$ equals to the bath temperature of the refrigerator $T_0$ due to the large substrate/bath contact area.

One of the devices used in the experiments is shown in Fig.~\ref {fig1}(b) and (c) together with the measurement setup. Cu film (brown) is evaporated on the silicon substrate, with 300~nm silicon oxide on top, by electron beam evaporation. The chamber pressure is kept below $5\times10^{-7}$ mbar during the deposition. Before contacting the Cu film with superconducting Al (blue), Ar plasma milling is used to clean the Cu film surface in order to achieve good metal-to-metal contacts between copper and aluminium. The hybrid structures with short channel length behave as a proximity Josephson junction (JJ). Switching current $I_{sw}$ is defined as the bias current when the junction switches from the superconducting state to the resistive state, shown in the IV curve in Fig.~\ref{fig1}(d). The JJ switches back to the superconducting state at a biasing current well below $I_{sw}$, defined as retrapping current $I_{r}$. The hysteresis of the IV curve originates from the overheating of the electrons after switching to the resistive state. Bath temperature dependence of $I_{sw}$ at zero heating, i.e., in equilibrium, shown in Fig.~\ref{fig1} (e), is used as the temperature calibration for the JJ thermometer \cite{Wang2018}. The long horizontal Cu wire between the large Cu pad and JJ thermometer is used as the heater to elevate electron temperature in the Cu film. We current-bias the two heater contacts with opposite polarities. Figure.~\ref{fig1} (f) is the measured $I_{sw}$ as a function of $I_{H}$ for various bath temperatures from 60~mK to 340~mK in 20~mK steps from top to bottom. Decrease of $I_{sw}$ while increasing $\mid I_{H}\mid$ indicates heating of the Cu film. The symmetry of the dependence around zero heating suggests no heating current flows to the thermometer in this configuration.

\begin{figure}[h]
	\centering
	\includegraphics[width=0.45\textwidth]{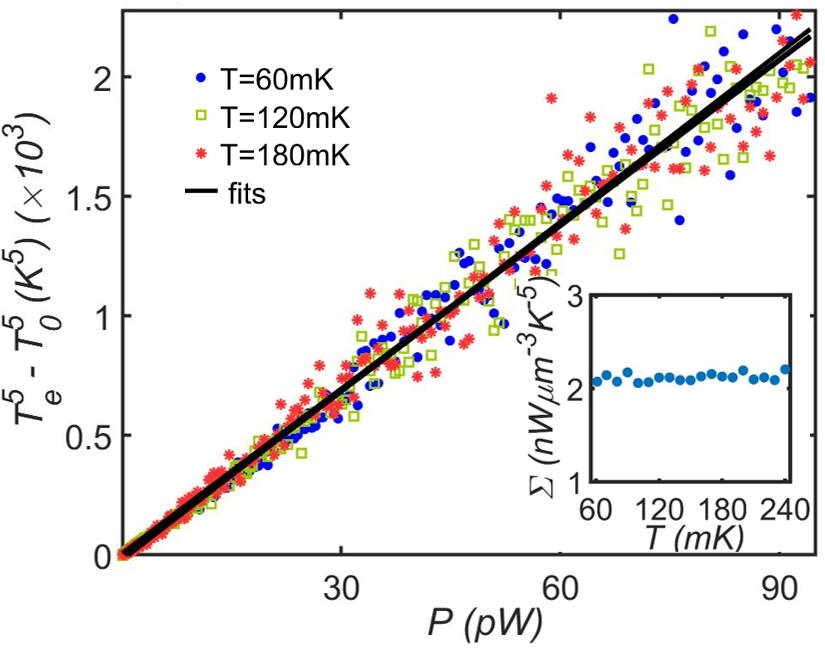}
	\caption{Measured $T_{e}^{5}-T_{0}^{5}$ plotted as a function of heating power $P$ for a sample with 50~nm Cu film. The observed linear dependence consistent with the prediction of Eq.~\eqref{eq1}, suggesting that the weak ep coupling limits the energy relaxation of the electrons in the Cu film. Inset is the measured ep coupling constant $\Sigma$ as a function of temperature.}
	\label{fig2}
\end{figure}

For electrons in the copper film, superconducting Al acts as a thermal insulator at sufficiently low temperatures below its critical temperature $T_c \sim 1$ K, the Joule power applied $P$ to the film dissipates mainly by ep coupling. The ratio of the two series thermal resistances is
\eq{
	\gamma = \frac{R_{pp}}{R_{ep}} = \frac{5\Sigma tT}{4k}. \label{eq3}
}
Here, $t$ is the thickness of the Cu film. For a thin film at sufficiently low temperatures, we expect $R_{ep}$ to dominate over $R_{pp}$, so we have the standard situation usually assumed for thin films, i.e., $P = P_{ep} = {\Sigma}V(T_{e}^{5}-T_{0}^{5})$. In Fig.~\ref{fig2}, we plot the experimental results of a sample with 50~nm thick Cu film, a linear dependence vs. $T_e^5 - T_0^5$ is clearly seen as expected. From the slope, we obtain the ep coupling constant $\Sigma$ = 2.1$\pm$0.1~nWK$^{-5}$$\mu$m$^{-3}$ with no temperature dependence within the measurement interval from 60~mK to 250~mK, as shown in the inset of Fig.~\ref{fig2}. The measured value of $\Sigma$ is consistent with previous experiments on Cu films \cite{Roukes1985,Meschke2004}. Thus, the experiment demonstrates that for the 50~nm Cu film at low temperature, the energy relaxation of electrons is dominated by the ep coupling, and the exponent $n = 5$ is consistent with the theory based on three-dimensional (3D) free electron model \cite{Wellstood1994}.

The Eq.~\ref{eq3} suggests that if one changes the film thickness or temperature to the point where $R_{pp}$ becomes equal to $R_{ep}$, a crossover from one energy relaxation mechanism to another should take place. The crossover temperature $T_{cr}$ depends on the constants $\Sigma$ and $k$ as $T_{cr}=4k/5\Sigma t$. For perfect contacts between Cu and the silicon substrate one finds $k \approx$ 170 WK$^{-4}$m$^{-2}$ \cite{Swartz1989}. Hence the crossover temperature of $0.1$~K is expected in films with the thickness of $t \approx 700$~nm. Recent experiments suggested that for evaporated films on a silicon substrate $k$ is smaller than that predicted for perfect contacts \cite{Rajauria2007,Pascal2013}, which makes it possible to observe $T_{cr}\approx 0.1$ K in somewhat thinner films.

\begin{figure}[t]
	\centering
	\includegraphics[width=0.5\textwidth]{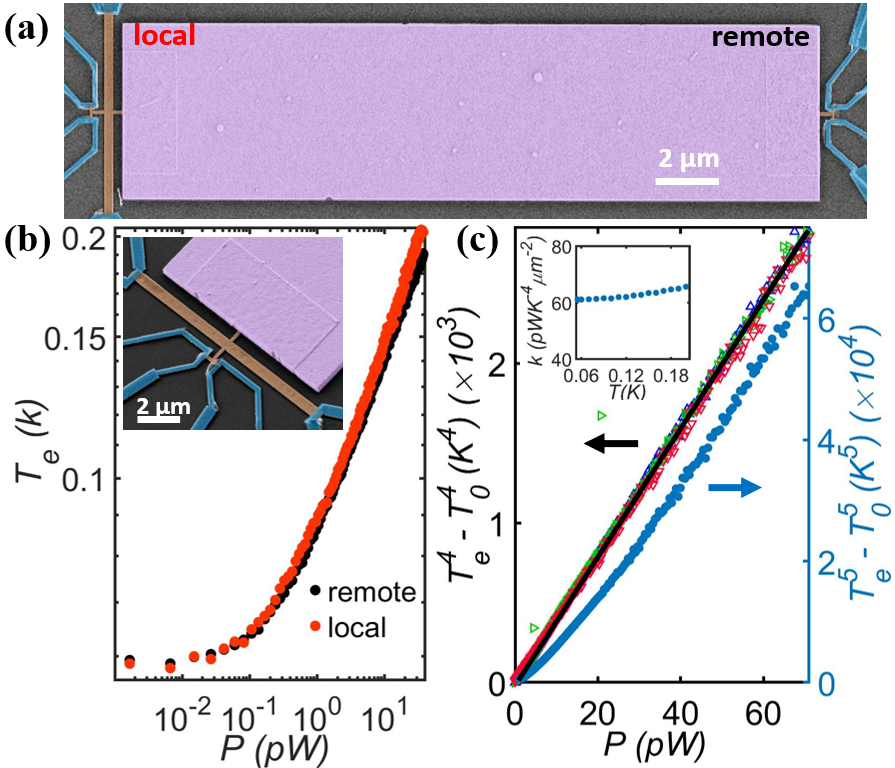}
\caption{(a) False-color SEM image of a sample with large Cu film (purple) of dimension 10~$\mu$m$ \times$ 40~$\mu$m$ \times$ 300~nm. Two JJ thermometers (local, remote) are located at the two ends of the Cu film to check the uniformity of the electron temperature while heating. Inset in (b) is the zoom-in of the local JJ thermometer showing the large Cu film covering the thin Cu film (brown). (b) Measured electron temperature by the local and the remote thermometer as a function of $P$, overlapping of the two curves suggesting electrons reach thermal equilibrium in the large Cu film. (c) Measured $T_{e}^{n}-T_{0}^{n}$ plotted as a function of $P$ with exponent $n$ = 4 (triangles, $T_0$  = 55~mK, 100~mK, 150~mK, from dark blue to red) and $n$ = 5 (blue dotted, $T_0$  = 55~mK). A linear dependence is observed when plotted with $n$ = 4, black lines are the linear fits. Experimental data show that for the 300~nm Cu film, thermal boundary resistance limits the energy relaxation process. The derived interface-material-dependent parameter $k$ is about 60 WK$^{-4}$m$^{-2}$, shown in the inset.} \label{fig3}
\end{figure}

In Fig.~\ref{fig3} (a), we show the SEM image of a sample with $t$ = 300~nm Cu film. Firstly, we deposit 50 nm Cu film (brown) used as JJ thermometers, heater, and the contact pads. Then we deposit the 300~nm Cu film (purple). Before contacting the two copper films, Ar plasma milling is used to clean the surface of the thin one. Inset of Fig.~\ref{fig3}(b) shows the thick film covering the thin film. Electron temperature is measured with two JJ thermometers located at the two ends of the thick Cu film (local, remote) with a distance of 40~$\mu$m to check the uniformity of electron temperature in the thick Cu film while heating. Figure.~\ref{fig3}(b) shows that the two thermometers show identical temperature except at the largest applied powers. The small difference at high $P$ originates most likely from the electron diffusion along the thick Cu film and is negligible for the analysis. The data also suggest that the thermal boundary resistance between the two Cu films is negligible.

We plot $T_{e}^{n}-T_{0}^{n}$ as function of $P$ in Fig.~\ref{fig3} (c). In contrast to what was seen in Fig.~\ref{fig2}, a linear dependence is observed when setting $n = 4$ in the full temperature range and three different bath temperatures explored. For comparison, we also show clearly non-linear dependence for $n$ = 5 and for the bath temperature $55$ mK with the blue dots. From the linear fit of $n$ = 4 data we have extracted the constant $k$ as a function of temperature, which is shown in the inset of Fig.~\ref{fig3} (c). 
We have found $k$ to be about 60 WK$^{-4}$m$^{-2}$ with a slight increase at high temperatures. The origin of this increase is unclear. 
The obtained value of $k$ is consistent with the previous experiments on evaporated metal films \cite{Rajauria2007,Pascal2013}, 
but it is smaller than the predictions of both AMM and DMM models. This difference may be explained by imperfect interface quality between the Cu film and the substrate.
\begin{figure}[h]
	\centering
	\includegraphics[width=0.5\textwidth]{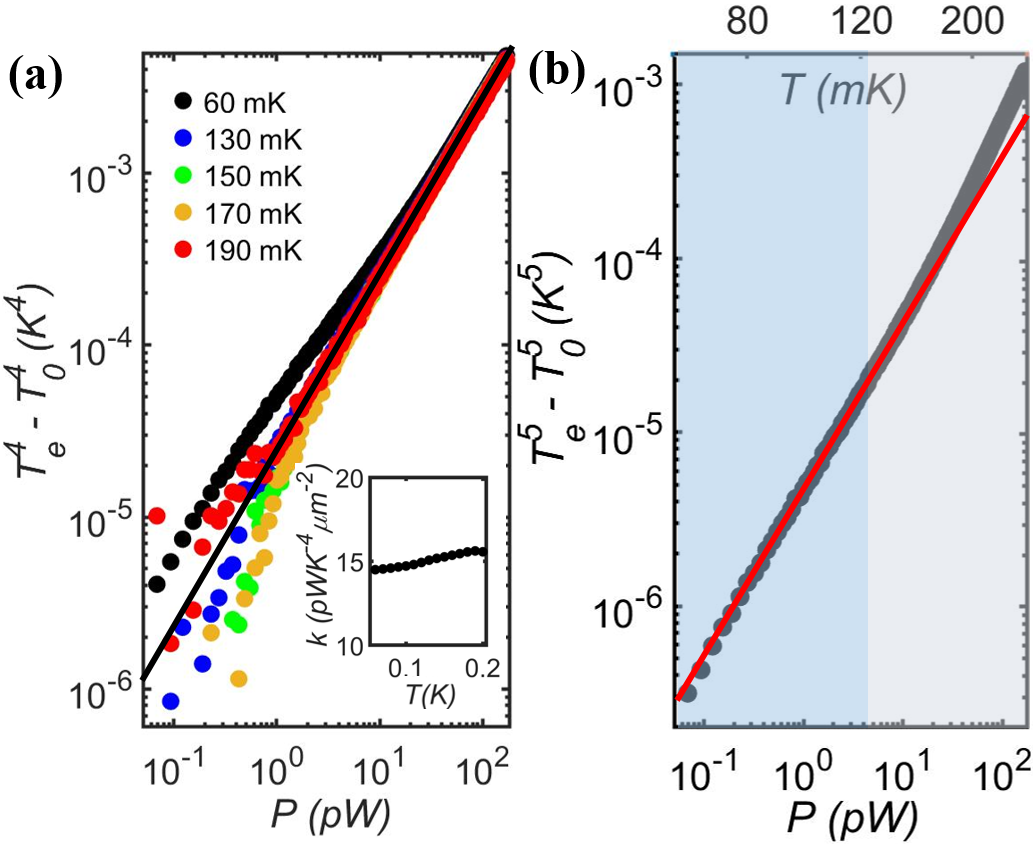}
	\caption{Measurement results of a sample with 3~nm Ti layer added between 50~nm Cu film and the substrate. (a)  $T_{e}^{n}-T_{0}^{n}$ as function of heating power $P$ with $n$ = 4. The observed linear dependence above 130~mK indicates that the thermal boundary resistance limits the energy relaxation. The black line is a guide to the eye. Inset: Derived $k$ as a function of temperature. (b) $T_{e}^{n}-T_{0}^{n}$ as a function of heating power $P$ with $n$ = 5 at 60~mK. Linear dependence is observed only at temperature up to about $T_{cr}$ = 120~mK. At higher temperatures, linear dependence is observed when plotted with $n$ = 4, shown in (a) with the black dotted line. The yellow line is a guide to the eye.}
	\label{fig4}
\end{figure}

Previous studies showed that for disordered normal metal films, the exponent $n$ deviates from 5 depending on the type of disorder \cite{Hsu1999,Sergeev2000,Gershenson2001,Karvonen2005}. The observed $n$ = 5 in 50~nm Cu film indicates the clean limit for the Cu film. Increasing the film thickness will reduce the disorder and make it closer to the 3D clean limit. So the observed $n$ = 4 for 300~nm Cu film is not to be ascribed to the film disorder. Instead, it originates from the fact that $R_{pp}$ dominates over $R_{ep}$ for thick film and becomes the bottleneck for heat transport.

As the acoustic mismatch between different materials will reduce the phonon transmission, an enhancement of the thermal boundary resistance is expected when adding a third layer of material between the Cu film and the substrate. We have fabricated a sample with 3~nm of Ti added between 50~nm Cu film and the substrate. By simply considering a series connection of the two interface resistances \cite{Dechaumphai2014,Li2012}, we find $k_{Cu-Ti-SiO_2}^{-1} = k_{Cu-Ti}^{-1} + k_{Ti-SiO_2}^{-1}$, where $k_{Cu-Ti}$ and $k_{Ti-SiO_2}$ are the interface-material-dependent parameters between Cu/Ti and Ti/SiO$_2$, respectively. 
In this way, from the DMM model we have estimated $k_{Cu-Ti-SiO_2}$= 0.38 $k_{Cu-SiO_2}$, where $k_{Cu-SiO_2}$ characterises the Cu/SiO$_2$ boundary.

In Fig.~\ref{fig4} (a), we plot measured $T_{e}^{4}-T_{0}^{4}$ as a function of $P$. Linear dependence is observed at temperatures above 130~mK, suggesting that $R_{pp}$ dominates the energy relaxation in this temperature range. From the linear fit we have estimated the constant $k_{Cu-Ti-SiO_2}$ to be about 15 WK$^{-4}$m$^{-2}$, 
which is 25\% of the value measured without the intermediate layer. It is a bit less than 38\% expected from the model discussed above. 
However, considering imperfect interface quality, and crudeness of the model, which, for example, ignores 
the fact that the thermal phonon wavelength is much larger than the thickness of the Ti film, 
our result is in a good agreement with the theory.
With the experimentally measured values of $k$ and ${\Sigma}$ we estimate the crossover temperature to be $T_{cr}$ = 124~mK. 
At temperatures below $T_{cr}$, $R_{ep}$ should dominate over $R_{pp}$ and the linear dependence of $T_{e}^{5}-T_{0}^{5}$ on $P$ is expected. In Fig.~\ref{fig4}(b), 
we show the measurement results at 60~mK. As expected, a linear dependence is observed at low temperatures, while for $T\gtrsim T_{cr}$ deviations from it become visible. 
In contrast, if one plots $T_{e}^{4}-T_{0}^{4}$ versus power, the linear dependence is observed at high temperatures $T\gtrsim T_{cr}$, as 
shown in Fig.~\ref{fig4}(a) with the black line. 
Thus, an additional 3~nm thin Ti layer between the Cu film and the substrate results in the fourfold increase in the thermal boundary resistance, 
which allows us to clearly see the crossover between the two energy relaxation mechanisms with changing temperature. 

One of the open questions is what the influence of the dimensionality of the film is on the acoustic coupling strength \cite{Kanskar1993,Nabity1991,Underwood2011}. It has been shown that even though the phonons in the film are 2D, the strong coupling of phonons in the film and the substrate can broaden its subband structure and make it closer to 3D. For Cu, the dominant phonon wavelength $\lambda$ is about 200~nm at 0.2~K when transverse phonons are considered, and it increases as $\propto T^{-1}$ when lowering the temperature. Assuming weak acoustic coupling and phonons in the film to be 2D, a reduction of the exponent $n$ from 5 is expected \cite{Glavin2002,Karvonen2007}. The observed $n$ = 5 for the 50~nm Cu film suggests phonons in the film are closer to 3D than 2D, though $\lambda$ is much large than the film thickness. The experimentally observed weaker acoustic coupling strength than what the theory predicts is not significant in making the phonons in the film 2D. Investigations are needed to quantify the effect of the strength of the coupling on the phonon dimensionality.

In conclusion, we have experimentally observed the crossover between the limiting energy relaxation mechanisms in copper films by changing the film thickness and temperature.
We have demonstrated that an additional Ti layer between the Cu film and the substrate enhances the thermal boundary resistance of the interface fourfold.
This result may be useful for hot-electron calorimetry and bolometry since it can help in improving the energy resolution of the detectors \cite{Pekola2013}. 
Our experimental results further advance the understanding of energy relaxation mechanisms in mesoscopic devices and of 
the heat transport through the solid-solid interfaces at low temperatures.

We acknowledge the provision of the fabrication facilities by Otaniemi research infrastructure for Micro and nanotechnologies (OtaNano). This work was performed as part of the Academy of Finland Centre of Excellence program (Projects No.312057.) and European Research Council (ERC) under the European Union's Horizon 2020 research and innovation program.(No. 742559 SQH).


\bibliography{TB2019}

\end{document}